\documentclass{jetpl}
\usepackage{bm}
\twocolumn

\lat
\DeclareMathOperator{\sgn}{sgn}

\title{Aharonov-Bohm oscillations caused by non-topological surface states in Dirac nanowires}

\rtitle{Aharonov-Bohm oscillations caused by non-topological surface states\ldots}

\sodtitle{Aharonov-Bohm oscillations caused by non-topological surface states in Dirac nanowires}

\author{V.\,V.\,Enaldiev$^a$, V.\,A.\,Volkov$^{a,b}$\thanks{e-mail: volkov.v.a@gmail.com, vova.enaldiev@gmail.com} }

\address{$^a$ Kotelnikov Institute of Radio-engineering and Electronics of the Russian Academy of Sciences, 11-7 Mokhovaya, Moscow, 125009 Russia}

\address{$^b$ Moscow Institute of Physics and Technology, 141700, Institutskii per. 9, Dolgoprudny, Moscow Region, Russia}

\dates{\today}{*}

\abstract{One intriguing fingerprint of surface states in topological insulators is the Aharonov-Bohm effect in magnetoconductivity of nanowires. We show that surface states in nanowires of Dirac materials (bismuth, bismuth antimony, and lead tin chalcogenides) being in non-topological phase, exhibit the same effect as amendment to magnetoconductivity of the bulk states. We consider a simple model of a cylindrical nanowire, which is described by the 3D Dirac equation with a general $T$-invariant boundary condition. The boundary condition is determined by a single phenomenological parameter whose sign defines topological-like and non-topological surface states. The non-topological surface states emerge outside the gap. In longitudinal magnetic field $B$ they lead to Aharonov-Bohm amendment for the density of states and correspondingly for conductivity of the nanowire. The phase of these magnetooscillations increases with $B$ from $\pi$ to $2\pi$. }

\begin{document}

\maketitle

{\bf 1. Introduction.}

In Dirac materials electron and hole energy spectra are described by a standard or modified Dirac equation (for review, see, for example \cite{Volkov_Enaldiev}). Recently introduced concept of topological characterization for bulk energy structure of crystals brings additional attention to such Dirac materials as bismuth, bismuth antimony alloys and lead tin chalcogenides \cite{Kane_Hasan, Qi_Zhang}. For definite percentage of antimony the alloy Bi$_{1-x}$Sb$_x$ becomes the topological insulator (TI), although pure bismuth is not the TI. The latter materials for certain content of tin \cite{Hsieh,Xu,Dziawa,Wojek}) are known as topological  crystalline insulators (TCIs). Surface states (SSs) in TIs and TCIs are protected by the time-inversion or mirror symmetry. The topological SSs result in the Aharonov-Bohm effect in magnetoconductivity of nanowire. In this paper we reveal that the Aharonov-Bohm effect can also exist in non-topological phase of the Dirac materials.

Lead tin chalcogenides have the rocksalt crystal structure with four inequivalent $L$-valleys in the reciprocal space. Bismuth antimony is of quite similar structure with an additional shear along a spatial diagonal of the cube, which reduces the number of $L$-valleys to the three ones. In the $L$-point of the Brillouin zone the both materials have two closely lying spin-degenerate energy bands in the vicinity of Fermi energy \cite{Dimmock, Wolff_PA}. Therefore minimal {\bf k$\cdot$p}-theory of the materials should include the two interacting bands. It leads to an anisotropic 3D Dirac equation that describes dynamics of fermions in the {\bf k$\cdot$p}-approximation. In this paper we will use a simplified isotropic version of the equation
\begin{equation}\label{Dirac_eq}
		\left\{ m\tau_z\otimes\sigma_0 +c\tau_x\otimes\left ( \bm{\sigma k } \right ) \right\}\Psi=E\Psi,
\end{equation}
 where $\bm{k}=(k_x,k_y,k_z)$ is the 3D momentum, $2m$ is the bulk energy gap between conduction and valence bands in the $L$-valley, $\bm{\sigma}=(\sigma_x,\sigma_y,\sigma_z) $ and $\bm{\tau} = (\tau_x,\tau_y,\tau_z)$ are vectors of the Pauli matrices acting in spin and band subspaces correspondingly, $\sigma_0$, $\tau_0$ are 2x2 identity matrices, $c$ is an effective speed of light, $\Psi = (\Psi_c,\Psi_v)$ is a four-component envelope function, $\Psi_c,\Psi_v$ are spinors describing conduction and valence bands respectively. 
 
Study of SSs for the 3D Dirac equation (\ref{Dirac_eq}) has a rather long history. First, Jackiw and Rebbi \cite{Jackiw_Rebbi} revealed that the Dirac equation has doubly degenerate bound states with zero energy if the mass term $m$ has opposite sign from two sides of an interface region. Later it was shown \cite{volkov_pankratov} that the interface states with 2D Weyl dispersion should arise in lead tin chalcogenides junctions where variation of tin content results in inversion of the mass term. One may use this treatment to phenomenologically consider TI \cite{Zhang_Kane} or TCI \cite{Liu_Duan} surface. 

Another approach to SSs of the Dirac equation (\ref{Dirac_eq}) is to derive an appropriate boundary condition (BC) for the four-component wave function $\Psi$ at a surface. This was done in Refs.\cite{volkov_pinsker} before the famous Ref.\cite{volkov_pankratov}, however the former paper has not been of much interest. The BC resulted from the Hermiticity of the Dirac Hamiltonian and time-reversal invariance of the problem in restricted area of space. Such a BC reads as follows
\begin{equation}\label{boundary_condition_main}
		\left ( \sigma_0\Psi_v-ia_0\bm{\sigma n}\Psi_c \right )_{\bm{r}\in S} = 0,
\end{equation}
where $\bm{n}=\bm{n}(S)$ is an inner normal to the surface $S$, $a_0$ is a real phenomenological parameter which characterizes atomic surface structure and bulk band properties. Afterwords \cite{Idlis_Usmanov} it was shown that to simulate the BC (\ref{boundary_condition_main}) via some interface potentials one should consider not only variation of the mass term but an additional electric potential which is responsible for electronic affinity of stacked materials.   

For special surface orientation we may additionally require the BC (\ref{boundary_condition_main}) to be invariant under mirror reflection if there is no surface reconstruction \cite{Zeljkovic}. For example, for $L$-valley, $[111]$ edge, of the rocksalt structure, in coordinate system $z||[111]$, $y||[1\overline{1}0]$, and $x||[\overline{1}\overline{1}2]$, $xz$-plane is a plane of mirror symmetry. If surface normal is in the $xz$-plane, i.e. $\bm{n}=(n_x,0,n_z)$, then invariance of the BC (\ref{boundary_condition_main}) under the mirror symmetry leads to the condition 
\begin{equation}\label{introduction_BC_invar}
\left. M_y\hat{\Gamma}\Psi\right |_{S}=\left. D(M_y)\hat{\Gamma}D^{-1}(M_y)D(M_y)\Psi\right |_{S} = \left. -\hat{\Gamma}\widetilde{\Psi}\right |_{S}=0,
\end{equation}
where the mirror symmetry operator $M_y$ is represented by the matrix $D(M_y)=-i\tau_z\otimes\sigma_y$, \mbox{$\widetilde{\Psi}=D(M_y)\Psi(x,-y,z)$}, the BC operator is 
\begin{equation}\label{isotropic_Gamma}
\hat{\Gamma}=\left ( 
\begin{array}{cc}
-ia_0\left ( n_x\sigma_x+n_z\sigma_z \right ) & 1 \\
0 & 0
\end{array}
\right ).
\end{equation}
Comparing (\ref{introduction_BC_invar}), (\ref{isotropic_Gamma}) with the BC (\ref{boundary_condition_main}) one can see that the mirror symmetry does not impose any restrictions on the value of the parameter $a_0$. 

\begin{figure*}
\begin{center}
			\includegraphics[width=15cm]{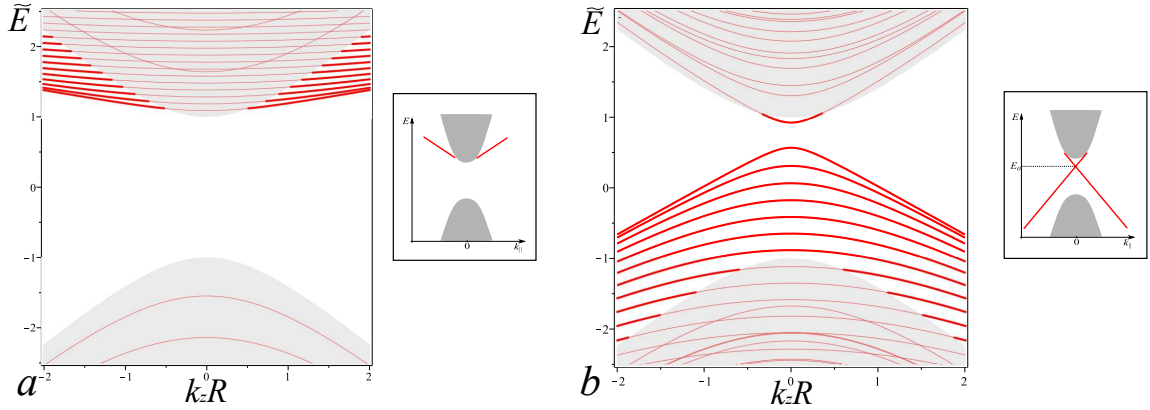} 
			\caption{ \label{Fig:QW_B=0} 
 Fig.1. Subband spectra ($\widetilde{E}=ER/\hbar c$) of Dirac fermions in a nanowire for two types of boundary parameter $a_0$: a) at $a_0=-0.1$ and $mR/\hbar c =2$ b) at $a_0=0.4$ and $mR/\hbar c=3$; bold red curves emphasize spectra of the non-topological surface subbands in Fig.\ref{Fig:QW_B=0}a and topological-like ones in Fig.\ref{Fig:QW_B=0}b. Spectra of SSs for half-space problem are shown on insets \cite{volkov_pinsker}. Grey colour fills regions of quantum-confined states. All subbands have double degeneracy due to the sign of $j$. }
\end{center}
\end{figure*}

In a halfspace $z\geq 0$ Eq. (\ref{Dirac_eq}) with the BCs (\ref{boundary_condition_main}) result in appearing of the 2D massless SSs with a conical spectrum \cite{volkov_pinsker} (see insets in Fig.\ref{Fig:QW_B=0})
\begin{equation}\label{dirac_spectrum}
E=sv|\bm{k}_{||}| + E_0,\quad  \frac{vm}{c^2\hbar} - s\frac{E_0}{m}|\bm{k}_{||}|  \geq 0,
\end{equation}
where \mbox{$v=2a_0c/(1+a_0^2)$} is the SS speed, \mbox{$\bm{k}_{||}=(k_x,k_y,0)$}, \mbox{$E_0=m(1-a_0^2)/(1+a_0^2)$} is the energy of the Dirac point, here $s=\pm 1$ are eigenvalues of a chirality operator $\tau_z\otimes (\bm{\sigma},[\bm{n},\bm{p}])$. Depending on the $a_0$ sign one can distinguish two qualitatively different cases: (i) $a_0\geq 0$ the SSs are in the bulk gap (red lines in inset in Fig.\ref{Fig:QW_B=0}b), and (ii) $a_0\leq 0$ the SSs are outside the gap (red lines in inset Fig.\ref{Fig:QW_B=0}a). Hence by means of the BC (\ref{boundary_condition_main}) one may phenomenologically describe the two types of SSs in lead tin chalcogenides existing in direct and inverse band order \cite{Khokhlov, Zeljkovic}.

Aim of our paper to study non-topological and topological-like SSs, for the sake of comparison, and to demonstrate their contribution in Aharonov-Bohm oscillations of the nanowire conductivity. We will suppose that $a_0$ is constant on the wire surface. Analogous approach was used to explain emergence and conducting nature of edge states circulating around a nanohole in graphene \cite{Latyshev}. Besides we calculate contribution of the SSs to the magnetoconductivity. It is of direct connection with recent experiments in magnetoresistance of bismuth \cite{Nikolaeva, Huber} and tin chalcogenide \cite{Safdar} nanowires. 

The paper is organized as follows. First, we calculate spectra of electrons in the nanowire without magnetic field. Second, we add a longitudinal magnetic field and also calculate contribution of the SSs to the magnetoconductivity.


{\bf 2. Dirac Fermions Spectra in Nanowire.}

Consider a cylindrical nanowire with radius $R$. We choose $z$-axis along the nanowire axis. The spinors $\Psi_c,\Psi_v$ obey the 3D Dirac equation (\ref{Dirac_eq}) and satisfy the BC (\ref{boundary_condition_main}) with the constant $a_0$ at a cylindrical surface of the nanowire. From now on we set $\hbar=c=1$ everywhere, except where it is needed. Cylindrical symmetry of the 3D Dirac equation (\ref{Dirac_eq}) with BCs (\ref{boundary_condition_main}) implies conservation of longitudinal momentum $k_z$ and total angular momentum $J_z=\sigma_0\otimes j_z$, with eigenvalues $j=\pm 1/2,\pm 3/2,\dots$, where $j_z=\sigma_0(-i\partial_{\theta})+\sigma_z/2$. Hence, one can find $\Psi_c$ as follows:
\begin{equation}\label{Psi_c}
\Psi_c=\left (
\begin{array}{c}
\psi_{c1}(r)e^{i(j-1/2)\theta} \\
\psi_{c2}(r)e^{i(j+1/2)\theta}
\end{array}
\right )e^{ik_zz}.
\end{equation} 
By use of the Dirac equation (\ref{Dirac_eq}), it is convenient to express $\Psi_v$ via $\Psi_c$. The radial wave functions $\psi_{c1}(r),\psi_{c2}(r)$ obey the Bessel equation 
\begin{equation}\label{radial_DE}
\begin{array}{l}
\left (-\frac{\partial^2}{\partial r^2}-\frac{\partial }{r\partial r} + \frac{(j\mp 1/2)^2}{r^2} \right )\psi_{c1,c2}=\left (E^2-m^2-k_z^2 \right )\psi_{c1,c2},
\end{array}
\end{equation}
and BCs
\begin{equation}\label{radial_BC}
\left.
\left ( 
\begin{array}{l}
i\left ( \partial_r - \frac {j-1/2}{R} -a_0(E+m)\right ) \quad k_z \\
-k_z \quad   i\left ( \partial_r + \frac {j+1/2}{R} -a_0(E+m)\right )
\end{array}
\right )
\left (
\begin{array}{l}
\psi_{c1}(r) \\
\psi_{c2}(r)
\end{array}
\right )
\right|_{r=R}=0.
\end{equation}
Therefore $\psi_{c1,c2}(r)=C_{1,2} J_{j\mp 1/2}(kr)$, where $k=~\sqrt{E^2-m^2-k_z^2}$, $J_{j\mp 1/2}(kr)$ are the Bessel functions of the first kind, $C_1,C_2$ are arbitrary constants.
The BCs (\ref{radial_BC}) impose a relationship for the constants $C_1,C_2$ and give the dispersion equation
\begin{equation}\label{QW_zeroB_DE}
\begin{array}{l}
k \left [ \dfrac{J_{j-1/2}(kR)}{J_{j+1/2}(kR)} - \dfrac{J_{j+1/2}(kR)}{J_{j-1/2}(kR)} \right ]= \quad\quad\quad\quad\quad\quad\quad\\
\\
\quad\quad\quad\quad\quad\quad\quad =\left ( a_0 - \dfrac{1}{a_0} \right )E + m\left ( a_0 + \dfrac{1}{a_0} \right ). 
\end{array}
\end{equation} 
Imagine values of $k$ correspond to the SSs, real values of $k$ describe quantum-confined states. Because of a symmetry property of the dispersion equation (\ref{QW_zeroB_DE}) $a_0\to 1/a_0, E\to -E$, it is enough to consider the case $|a_0|\leq 1$. Besides, the dispersion equation does not depend on $j$ sign as follows from the property of the Bessel function of integer index \mbox{$J_{-j\mp 1/2}(x)=(-1)^{j\pm 1/2} J_{j\pm 1/2}(x)$}. It reflects the axial symmetry of the nanowire. Consequently all 1D subbands have double degeneracy due to the $j$ sign.  Result of qualitative graphic solution of the dispersion equation (\ref{QW_zeroB_DE}) is shown on Fig.(\ref{Fig:QW_B=0}). 
The energy of quantum-confined states at $|k_z|\to~\infty$ aspires to $E=\pm\hbar c\sqrt{m^2/\hbar^2c^2 + k_z^2 + \gamma_{j\pm 1/2,n}^2/R^2}$, where $\gamma_{j\pm 1/2,n}$ is the n-th zero of the Bessel function $J_{j\pm 1/2}$. At $a_0>0$ the SSs acquire a mass equals $\hbar v|j|/R$ in the thick wire limit ($|a_0| mR/\hbar c\gg 1$). The energy spectrum of the SSs has asymptotes 
\begin{equation}\label{QW_SS_Dispersion}
\begin{array}{l}
E=sv\hbar\sqrt{k_z^2+\frac{j^2}{R^2}}+E_0, \\ \kappa=\frac{vm}{c^2\hbar} - s\frac{E_0}{m}\sqrt{k_z^2+\frac{j^2}{R^2}}\geq 0
\end{array}
\end{equation} 
when their decay length $\kappa^{-1}$ is much smaller than the radius $R$ $(\kappa R\gg j^2) $. The closer energy of the SSs to the region of quantum-confined states (this region is filled with grey color on the Fig.\ref{Fig:QW_B=0}) the more delocalized are the wave functions of the SSs.


{\bf 3. Dirac Electrons in nanowire in longitudinal magnetic field: Spectra and Conductivity.} 

\begin{figure*}
			\includegraphics[width=16cm]{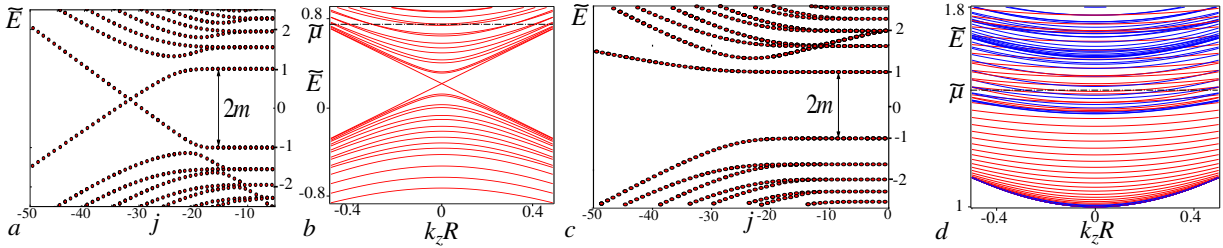} 
			\caption{ \label{Fig:QW_in_B_E(j)} 
Fig.2. Energy spectrum of Dirac fermions in a nanowire with a longitudinal magnetic field for two types of boundary parameter $a_0$: a,c) Dimensionless spectra $\widetilde{E}=ER/\hbar c$ of 1D subband bottoms as a function of total angular momentum $j$ for $a_0=0.8$ (a) and $a_0 =-0.2$ (c). b,d) Spectra of 1D subbands for $a_0=0.8$, $j=[-29.5, -21.5]$ (b) and $a_0 =-0.2$ $j=[-51.5, -7.5]$ (d). On Fig. \ref{Fig:QW_in_B_E(j)}d red curves correspond to the non-topological surface subbands, blue curves -- Landau and skipping orbit  subbands. Numerical solution was carried out at $\hbar c/\lambda=0.896m$, $\Phi/\Phi_0=35$. Dashed-dotted straight line on figures b) and d) indicates the Fermi-level $\widetilde{\mu}$.
  }
\end{figure*}

In magnetic field along the wire axis $\bm{B}=(0,0,B)$ we make the Peierls substitution $\bm{p}\to\bm{p}+e\bm{A}$ in the 3D Dirac equation (\ref{Dirac_eq}), $-e$ is the electron charge. For the vector-potential we choose the cylindrical gauge $\bm{A}=~(-By/2,Bx/2,0)$. Further, to be specific we consider the case $B>0$. The radial components $\psi_{c1},\psi_{c2}$ of the spinor $\Psi_c$ obey the following equation
\begin{equation}\label{radial_DE_in_B}
\begin{array}{l}
	\left (-\dfrac{\partial^2}{\partial r^2} - \dfrac{\partial}{r\partial r} + \dfrac{(j\mp 1/2)^2}{r^2} + \dfrac{j\pm 1/2}{\lambda^2}+\dfrac{r^2}{4\lambda^4} \right )\psi_{c1,c2}= \\
	\\ 
	\qquad\qquad(E^2-m^2-k_z^2)\psi_{c1,c2}, 
\end{array}
\end{equation}
where $\lambda^2=1/eB$ is the magnetic length squared. Normalizable solutions of the radial equations (\ref{radial_DE_in_B}) are expressed in terms of the Kummer's function $M(\alpha,\beta,\xi)$ \cite{Abramovic}. While $B>0$ we are interested in spectra of SSs with $j\leq -1/2$. The dispersion equation for those $j$ is: 
\begin{equation}\label{DE_in_B}
\begin{array}{l}
\left [ 2(j-1/2) - a_0R(E+m)\widetilde{M}\right ]\times\\
\quad\left [\dfrac{R^2k^2}{2(j-1/2)}+\dfrac{a_0R(E+m)}{\widetilde{M}} \right] + k_z^2R^2=0.
\end{array}
\end{equation}
where 
\begin{equation} \widetilde{M}=\dfrac{M(1-\lambda^2k^2/2,-j+3/2,R^2/2\lambda^2)}{M(-\lambda^2k^2/2,-j+1/2,R^2/2\lambda^2)}\nonumber.
\end{equation}
Numerical solution of the equation (\ref{DE_in_B}) yields the energy spectrum of the Dirac fermions in the nanowire in the longitudinal magnetic field, see Fig. \ref{Fig:QW_in_B_E(j)}.
In two limiting cases the dispersion law of the SSs is: 
\begin{eqnarray}\label{QW_SS_Dispersion_in_B}
E_{k_zjs}=sv\hbar\sqrt{k_z^2+\dfrac{\left ( j+\Phi-\gamma_B\right )^2}{R^2}} + E_0,  \nonumber \\
\end{eqnarray}
where $\Phi=\pi eBR^2/hc$ is the number of the magnetic flux quanta through the wire cross section, with $\gamma_B = 0$ in a quasiclassical limit $|\kappa R/j|\gg \max{|j|,\Phi}$ (weak magnetic field and strong localization for radial motion
of the SSs), and $\gamma_B=1/2$ in the limit of strong magnetic fields ($\Phi\gg|j-1/2|$, $\Phi\gg \lambda^2k^2/2$, the last condition results in $|a_0+1/a_0|\gg 1$ in the Eq.(\ref{QW_SS_Dispersion_in_B})). For $a_0>0$ both values of $s=\pm 1$ are valid in Eqs.(\ref{QW_SS_Dispersion_in_B}) and surface subbands are formed in the bulk gap. In this case situation resembles that of topological insulator nanowire \cite{Bardarson_Moore, Zhang_Ran, Ioselevich_Fegelman}. In the case $a_0<0$ only one value of $s$ is allowed and surface subbands coexist with Landau and skipping orbit subbands (see Fig.\ref{Fig:QW_in_B_E(j)}c).    

While the magnetic field is relatively weak, the gapless SS subbands periodically emerge at half-integer values of $j_0\equiv-\Phi$ with unit period. With further increase of the magnetic field when the magnetic length compared with the SS decay length $\kappa^{-1}$, the period is of weakly dependence on $B$. Far from Dirac point $(E-E_0)R/\hbar v\equiv pR \gg 1 $ 

Density of SSs with specta (\ref{QW_SS_Dispersion_in_B}) is described by the formula: 
\begin{eqnarray}\label{density of states}
\rho(E) = \rho_0\left ( 1 + 2\sqrt{\frac{2}{\pi}}S\left(E\right)\right)\times\nonumber \\
\times\Theta\left [ \left (E-E_0\right)\sgn(a_0(a_0^2-1))+\hbar vk_e\right ], 
\end{eqnarray} 
where $\Theta[...]$ is the Heaviside step function which determines energy region where SSs exist ((for $0<a_0\leq 1$ the region is $E<E_0+\hbar vk_e$, for $a_0 > 1$ the region is $E > E_0 -\hbar vk_e$, for $-1 < a_0 < 0$ the region is $E >
E_0-\hbar v k_e$, and for $a_0\leq 1$ the region is $E < E_0+ \hbar vk_e$)), $\rho_0=(E-E_0)/2\pi(\hbar v)^2$ is density of states (DOS) of free massless 2D DFs, $k_e=2|a_0|m/\hbar c|1-a_0^2|$, $S$ is oscillating amendment to free DOS due to quantization of perimetric motion of SSs:
\begin{equation}\label{DOS_amendment}
S\left(E \right) = \sum\limits_{n=1}^{+\infty}\frac{\cos\left( 2\pi pRn-\frac{\pi}{4}\right )\cos(2\pi n \left(\Phi  - \frac{1}{2} - \gamma_B\right) )}{\sqrt{2\pi pRn}}
\end{equation}
When energy in (\ref{DOS_amendment}) coincides with a bottom of every
surface subband the quantities $pR\pm(\Phi-1/2 + \gamma_B)$ tend to integers $N_{\pm}$. In this case the series (\ref{DOS_amendment}) can be estimated
as (see \cite{Ioselevich}):
\begin{eqnarray}\label{series_estim}
S(\mu ) 
\approx \frac{\Theta\left(pR\pm\left(\Phi-\frac{1}{2}-\gamma_B\right) - N_{\pm}\right )}{2\sqrt{2\pi pR\left(pR\pm\left(\Phi-\frac{1}{2}-\gamma_B\right) - N_{\pm}\right)}},
\end{eqnarray}
It is seen that in this case $S$ has square root singularities as it should be for DOS of one-dimensional subbands.

{\bf 4. Conductivity of Surface States.}
Let us calculate the electric conductivity $\sigma$ of the $\ell$-length nanowire in static uniform electric field of strength $F$. At first we consider the case $a_0>0$ when Fermi energy $\mu$ is in the bulk gap (see Fig.\ref{Fig:QW_in_B_E(j)}b). In this case only the SSs contribute to the conductivity at low temperatures. We neglect by intervalley scattering and consider weak scalar disorder potential as follows $V(\bm{r}) = \sum_{i=1}^{N}u(\bm{r}-\bm{r}_i)$ with $u(\bm{r}-\bm{r}_i) =U\delta(r-r_i)\delta(\theta-\theta_i)\delta(z-z_i)/r_i$. We will employ wave functions of SSs in the quasiclassical limit: 
\begin{equation}\label{SS_wave_function}
\begin{array}{l}
|j,k_z,s\rangle\approx  \\
C_{k_zj}\left (
\begin{array}{c}
e^{i(j-1/2)\theta} \\
-\dfrac{is}{k_z}\sqrt{k_z^2+\frac{(j+\Phi)^2}{R^2}} e^{i(j+1/2)\theta}\\
\dfrac{sa_0}{k_z}\sqrt{k_z^2+\frac{(j+\Phi)^2}{R^2}} e^{i(j-1/2)\theta}\\
ia_0 e^{i(j+1/2)\theta}
\end{array}
\right )\dfrac{e^{\kappa(r-R)+ik_zz}}{\sqrt{2\pi\kappa r}}.
\end{array}
\end{equation}
Here we suppose that $\mu$ lies well above $E_0$ (but in the bulk gap) so that a lot of surface subbands with $s=1$ to be filled ($QR\gg 1$ with $\hbar v Q=\mu - E_0$). In case of weak impurities while the characteristic time $\tau$ of change of the electron's distribution function much greater than the period of the evolution operator oscillation ($\sim \hbar/\mu$) we can restrict ourselves by $\tau$-approximation in the Boltzmann kinetic equation. We also suppose that one-dimensional localization length is much greater than the nanowire length. It is easy to meet this condition while the number of filled surface subbands is great ($QR\gg 1$). Therefore one may assume the specific 1D localization effects to be small. In the given approximation the current in $z$ direction is expressed by the formula \cite{Sandom} 
\begin{equation}\label{current}
j_z = -\frac{g_ve^2F}{\ell}\sum_{\nu}\left (v_z\right )_{\nu}^2\tau_{\nu}\left ( \left. -\frac{\partial f_0}{\partial E}\right |_{E=E_{\nu}} \right ),
\end{equation}
here $\nu = \left\{k_z,j \right\}$, $\left (v_z \right )_{\nu}$ is the diagonal matrix element of the velocity operator of the Dirac Hamiltonian, $g_v$ is the number of non-equivalent $L$-valleys in the Brillouin zone, $\tau_{\nu}$ is relaxation time in the state $|\nu\rangle$ that expressed as follows 
\begin{equation}\label{relaxation_time}
\tau^{-1}_{\nu} = \sum_{\nu'}  W_{\nu\nu'}\left ( 1-\dfrac{\left(v_z\right )_{\nu'}}{\left(v_z\right )_{\nu}} \right ).
\end{equation}
In the above formula transition probability $W_{\nu\nu'}$ averaged over configuration of scatterers is expressed by the following way
\begin{equation}\label{transition_prob}
\begin{array}{l}
W_{\nu\nu'} = \\ \frac{2\pi}{\hbar}\sum_{i}\int_{0}^{R}\frac{2r_idr_i}{R^2}\left | \langle \nu| u_{\nu-\nu'}(r)e^{i(j-j')\theta+i(k_z-k_z')z}|\nu' \rangle \right |^2\times \\
\delta(E_{\nu} - E_{\nu'}), 
\end{array}
\end{equation}
where we introduce the Fourier coefficient of $u(\bm{r}-\bm{r}_i)$ 
\begin{equation}
\begin{array}{l}
u_{\lambda}(r) = \frac{1}{2\pi\ell}\int_{-\ell/2}^{\ell/2}\int_{0}^{2\pi}u(\bm{r}-\bm{r}_i)e^{-im(\theta-\theta_i)-iq(z-z_i)}dzd\theta = \\
= \frac{U}{2\pi\ell r_i}\delta(r-r_i), \quad\lambda=\left\{ m,q\right\}.
\end{array}
\end{equation}
Using the Poisson summation formula one can obtain the following equation for relaxation time (\ref{relaxation_time}) in the leading order in $(pR)^{-1/2}$: 
\begin{equation}\label{relax-time-final}
\begin{array}{l}
\tau^{-1}_{\nu} =
\frac{\pi n_{im}}{\hbar}\frac{U^2\kappa}{R} \rho_0(E_{\nu})\frac{ \left (2-\sqrt{2}+ \frac{\tilde{j}^2}{\sqrt{2}k_z^2R^2} \right )}{\left(1+\frac{\tilde{j}^2}{2k_z^2R^2} \right )}\times
\\
\left [ 1 + \frac{4\sqrt{2}\left ( 1 + \frac{\tilde{j}^2}{k_z^2R^2} \right )}{\sqrt{\pi}\left ( 2 - \sqrt{2} + \frac{\tilde{j}^2}{\sqrt{2}k_z^2R^2} \right )}S(E)\right ],
\end{array}
\end{equation} 
where $n_{im}=N/2\pi R\ell$ is 2D concentration of impurities, $\tilde{j} = j+\Phi-\gamma_B$. Only impurities, that are located in layer of $\kappa^{-1}$ thickness around the surface of nanowire, effectively scatter the Dirac fermions in SSs. Additional phase shifting $\pi n$ in the last term in (\ref{relax-time-final}) is manifestation of the Berry phase of the SSs. Under our consideration ($pR\gg 1$) the sum in the above equation is a small oscillating amendment to non-quantized SS density of states $\rho_0$. It is the amendment that leads to the Aharonov-Bohm oscillations of observable quantities. To obtain conductivity of the nanowire we substitute relaxation time from Eq. (\ref{relax-time-final}) into Eq. (\ref{current}). After some cumbersome integration we come to the final result in leading approximation in $(QR)^{-1/2}$:
\begin{equation}\label{conductivity}
\sigma = \sigma_{2D}2\pi R 
\left [ c_1 - c_2S(\mu) \right ]
\end{equation}
where $c_1\approx 0.95$, $c_2\approx 10.6$, and
\begin{equation}
\sigma_{2D} = \frac{e^2v^2\hbar R}{\pi n_{imp}U^2\kappa}
\end{equation}
is conventional conductivity of 2D massless DFs with short-range impurities with corresponding replacement of potential strength $U^2\kappa/R\to u_0^2$ \cite{Hong_Ando}. Series in Eq.(\ref{conductivity}) converges for every value of magnetic flux $\Phi$ but the fluxes when Fermi level coincides with bottoms of the surface subbands. As it was mentioned in the previous section this occurs at integers $QR\pm(\Phi- 1/2 - \gamma_B)$. In this case the $\tau$-approximation is not valid even for weak impurities. Eq.(\ref{conductivity}) is valid only when the Fermi level is quite far from surface subband bottoms where $S\ll 1$.

For the case $a_0<0$ similar calculation can be performed taking only transitions between the SSs. This contribution to the conductivity is of the form:
\begin{equation}\label{conductivity_a_0<0}
\delta\sigma = \sigma_{2D}2\pi R\left[ c_1-c_2S(\mu )\right]\Theta\left[\left(\mu-E_0 \right )\sgn(1-a_0^2) + \hbar vk_e\right].
\end{equation}
Heaviside step function in above equation means that this contribution is not zero when Fermi level lies in conduction or valence band, depending on the value of $a_0$, and intersects surface subbands. 

Now, we discuss situation when the Fermi level lies in the vicinity of the Dirac point $E_0$  at $a_0\geq 0$. Two gapless subbands arise at half integers $\Phi$. Backscattering between the gapless states moving in opposite directions is forbidden as $W_{\nu\nu'}$ is zero in Eq.(\ref{transition_prob}).  Therefore, in this regime conductance of the nanowire will $\Phi_0$-periodically oscillate with maximum value $g_ve^2/h$ at half when massless subband arise \cite{Bardarson_Moore}. (in weak magnetic field at half integers $\Phi$, but in strong magnetic field limit at integers $\Phi$ (see Eq.(\ref{QW_SS_Dispersion_in_B})))


{\bf Conclusions.} We reveal that the $T$-invariant boundary condition for 3D Dirac equation in nanowire can lead to emergence of topological-like or non-topological types of SSs depending on sign of the BC parameter $a_0$. This parameter is determined by microscopic surface structure and bulk band properties. At $a_0>0$ topological-like SSs forming in the bulk gap as in topological insulator nanowires \cite{Bardarson_Moore, Zhang_Ran,Ostrovskiy,Ioselevich_Fegelman} may lead to the Aharonov-Bohm oscillations of magnetoconductivity. Although at $a_0 < 0$ non-topological surface subbands emerge out of the gap, they also may lead to the Aharonov-Bohm oscillations of the nanowire conductivity. The phase of magnetooscillations is $\pi$ for both types of surface subbands (i.e. both
signs of $a_0$) in quasiclassical limit, but it acquires additional phase shift $\pi$ in strong magnetic field limit. Thus, observation of the Aharonov-Bohm conductivity magnetooscillations (or surface contribution in zero $B$) in the Dirac nanowires is not evidence for topological nature of the SSs.   


{\bf Acknowledgements.} This work was supported by the Russian Science Foundation (grant 16-12-10411). 


\end{document}